# EFFECT OF POST WELD HEAT TREATMENTS ON THE ELEVATED TEMPERATURE MECHANICAL PROPERTIES OF Ti6Al4V FRICTION WELDS

**RAHUL[1#], K. V. RAJULAPATI[1], G. M. REDDY[2], T. MOHANDAS[3], K. B. S. RAO[4]**

[1]School of Engineering Sciences and Technology, University of Hyderabad, Hyderabad-500046, India
[2]Defence Metallurgical Research Laboratory, Hyderabad-500058, India
[3]Nalla Malla Reddy Engineering College, Hyderabad-500088, India
[4]Ministy of Steel (Govt. of India) Chair Professor, Mahatma Gandhi Institute of Technology, Hyderabad-500075, India

[#]luhar6100@gmail.com

## ABSTRACT

*The α+β titanium alloy (Ti6Al4V) has been successfully joined using rotary friction welding. To investigate the influence of post weld heat treatments on the microstructure and mechanical properties of the welds, the weld joints were heat treated in α+β and β regions, followed by air cooling and furnace cooling. Subsequent to heat treatment, the specimens were subjected to stress relieving treatment. The heat treatment temperatures were selected keeping in view the beta transus temperature of the alloy. Mechanical properties of the welds are evaluated in the as-welded and post weld heat treated conditions at the working temperature of this alloy. The results reported are an average of the values obtained from three tests carried out at a given set of condition. Joints produced exhibited better mechanical properties when compared to the parent metal. Coarse transgranular microstructure and coarse grains in general exhibit better creep and stress rupture properties, while finer microstructures exhibit better tensile strengths.*

## INTRODUCTION

Titanium alloys are being widely used in aerospace industry, particularly for compressor stage of gas turbine engine. The advantages of using titanium are its outstanding specific strength compared to aluminium and steel over a wide range of temperatures, excellent elevated temperature properties and corrosion resistance [1]. The occurrence of an allotropic modification of titanium from the high temperature β – phase (BCC) to the room temperature α – phase (HCP), the strong dependence of the beta transus temperature on the alloy composition and the variety of phase transformations in titanium alloy systems are some of the factors that allow wide microstructural control, such that the properties can be tailored to specific needs [2].

Titanium has high affinity towards atmospheric gases such as oxygen and nitrogen.

Excessive amounts of these gases in titanium cause embrittlement. In addition to this, fusion welds are prone to solidification related defects resulting in loss of mechanical properties [3]. In order to overcome the problems related to fusion welding, attempts have been made to introduce solid state joining processes like friction welding.

In addition, the base metal microstructure has been reported to have a significant influence on the fusion zone grain size of the alloy particularly when the welding is carried out autogenously as in the case of electron beam welding [4]. As a consequence, it has been reported that the mechanical properties and fracture characteristics will differ depending upon the microstructure [5].

## EXPERIMENTAL DETAILS

Parent metal employed in this study was Ti6Al4V alloy in 16 mm diameter rod form, in the hot rolled and mill annealed condition, supplied by Mishra Dhatu Nigam (MIDHANI) Limited, Hyderabad, India. Microstructural characterization was carried out using Hitachi S3400N scanning electron microscope (SEM). Heat treatments were carried out in a box type muffle furnace. Specimens were coated with delta-glaze prior to heat treatment, inorder to protect the specimens from oxidation. α+β heat treatment condition consisted of heating to a temperature of about 950 $^{o}$C and holding for 1 h followed by air cooling (AC) and furnace cooling (FC). Similarly β heat treatment condition consisted of heating to a temperature of about 1020 $^{o}$C and holding for 20 min followed by air cooling and furnace cooling. The heat treated specimens were subjected to stress relieving treatment at a temperature of about 540 $^{o}$C for 4 h followed by air cooling. Description of the designations given to the samples is presented in Table 1.

A 150 kN capacity continuous drive friction welding machine (ETA Technology make, Bangalore, India) was used to carry out friction welding. Welding was carried out at 1500 rpm, 3 kN friction force, 5 kN upset force and 7 s friction time. Tensile tests were carried out at a cross head speed of 1 mm/min, which results in an initial strain rate of $0.55 \times 10^{-3} s^{-1}$. Creep tests were performed under an applied stress of 320 MPa for a period of 100 h. Stress rupture properties were evaluated at a stress level of 520 MPa, until failure. All the tests were carried out at a temperature of 400 $^{o}$C.

Table 1. Nomenclature of the specimens employed in the present study.

| Specimen ID | Description |
| --- | --- |
| ASR | As-received parent metal |
| FRW | As-welded parent metal |
| ASR BB AC | ASR + α+β heat treatment + AC + stress relieving |
| FRW BB AC | FRW + α+β heat treatment + AC + stress relieving |
| ASR AB AC | ASR + β heat treatment + AC + stress relieving |
| FRW AB AC | FRW + β heat treatment + AC + stress relieving |
| ASR BB FC | ASR + α+β heat treatment + FC + stress relieving |
| FRW BB FC | FRW + α+β heat treatment + FC + stress relieving |
| ASR AB FC | ASR + β heat treatment + FC + stress relieving |
| FRW AB FC | FRW + β heat treatment + FC + stress relieving |

## RESULTS AND DISCUSSION

Parent metal in the as-received condition, i.e., mill-annealed condition, contained a microstructure of elongated equiaxed α grains with discontinuous β phase at the grain boundaries (Figure 1a). In the SEM micrographs, the white region is β phase and the darker region is α phase. At higher magnification it was observed that a series of discontinuous β phase formed at the grain boundary of α grains. It is also observed that the grains are elongated in the rolling direction, and the structure is un-recrystallized.

The parent metal when subjected to heat treatment in the α+β region, followed by air cooling, resulted in a microstructure containing equiaxed primary α grains in a transformed β matrix (Figure 1b). β heat treatment of the parent metal followed by air cooling exhibited a microstructure that contained coarse prior β grains with colonies of coarse α platelets and a thin layer of β between the α platelets (Figure 1c). Furnace cooling of the parent metal subjected to heat treatment in the α+β region resulted in an equiaxed α structure with intergranular discontinuous network of the β phase (Figure 1d). β heat treatment of the parent metal followed by furnace cooling resulted in grain coarsening. Widmanstatten structure or basket weave structure is observed to have formed within the α grains (Figure 1e).

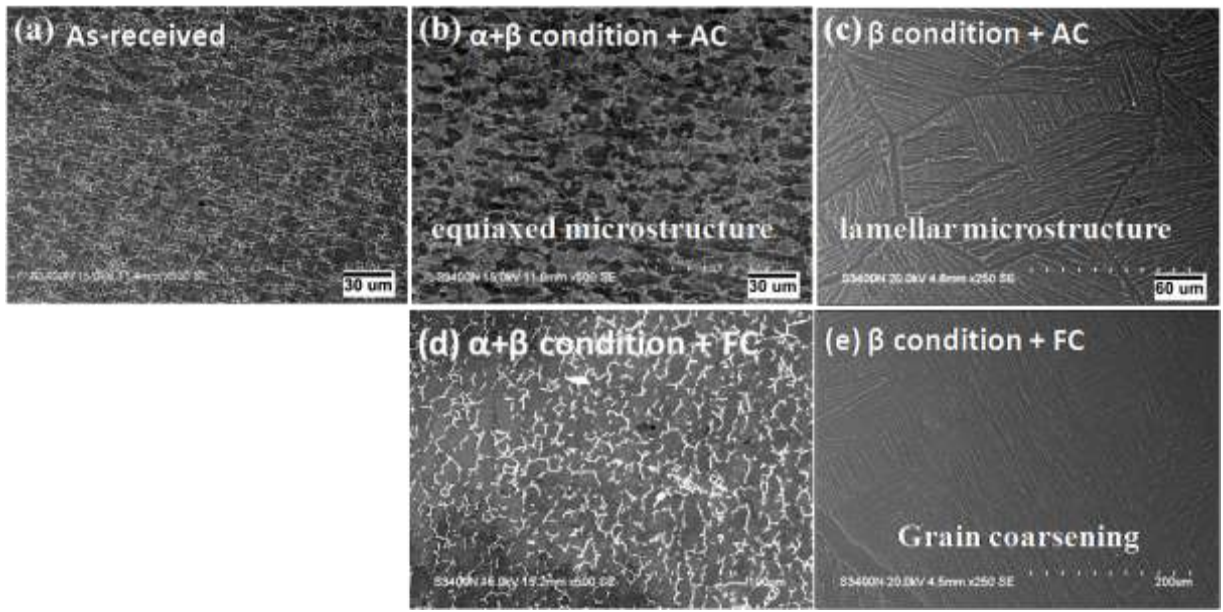

Figure 1. Microstructures of the Ti6Al4V parent metal in the as-received condition, as well as in the heat treated conditions.

Examination of the weld region using scanning election microscope showed that the weld joints in the as-welded condition, contained acicular product of α$^/$+β with random orientation (Figure 2a). Post weld heat treatment of the welds in the α+β region followed by air cooling resulted in coarsening of the acicular product with thickening of α phase. The structure mainly contained colonies of alternate layers of α and β with α along prior β grain boundaries (Figure 2b). Furnace cooling further coarsened the structure leading to the formation of thick and high volume fraction of α, and randomly oriented, randomly distributed thin α as acicular product (Figure 2d). Microstructure of the welds, after heat treatment in the β region, transformed completely into lath α+β, in prior β grains as a result of air cooling (Figure 2c) and plate-like α+β in the prior β grains as a result of furnace cooling (Figure 2e). The structure of the weld completely transformed after β heat treatment and is same as the parent metal and the weld joint could not be distinguished.

The variations in microstructures observed as a result of heat treatments are in conformity with those reported in literature [6-9].

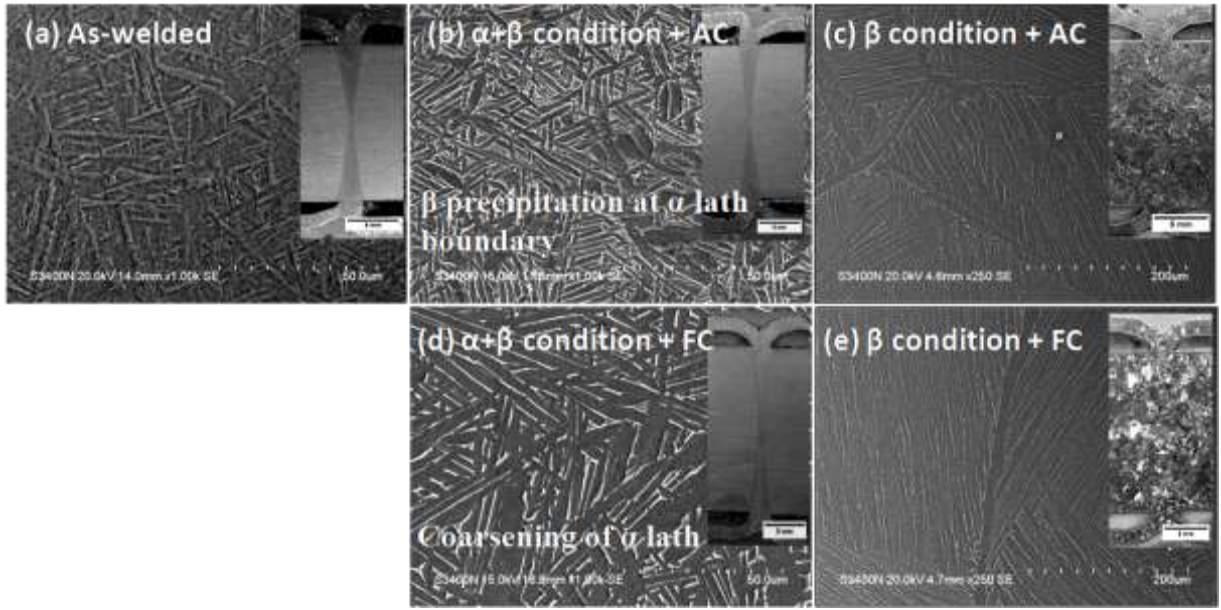

Figure 2. Microstructures of the Ti6Al4V friction welded joint in the as-welded condition, as well as in the post weld heat treated conditions.

Table 2. Mechanical properties of Ti6Al4V parent metal and welds under various heat treatment conditions.

| Specimen ID | Tensile Yield Strength (Mpa) | Ultimate Tensile Strength (Mpa) | Elongation (%) | Time to Rupture (h) | Creep strain after 100h (%) |
|---|---|---|---|---|---|
| ASR | 557 ± 7 | 653 ± 3 | **21** | 194 | 0.42 |
| FRW | **563 ± 2** | **679 ± 5** | 18.3 | **212** | 0.39 |
| ASR BB AC | 509 ± 5 | 672 ± 9 | 19 | 129 | 0.53 |
| FRW BB AC | 511 ± 7 | 669 ± 3 | 18 | 144 | 0.21 |
| ASR AB AC | 560 ± 4 | 675 ± 2 | 9.3 | 158 | 0.38 |
| FRW AB AC | 554 ± 4 | 670 ± 1 | 8.7 | 171 | 0.37 |
| ASR BB FC | 482 ± 3 | 613 ± 4 | 20.9 | 77 | 0.67 |
| FRW BB FC | 484 ± 3 | 604 ± 1 | 18.6 | 104 | 0.34 |
| ASR AB FC | 523 ± 2 | 621 ± 9 | 11.7 | 30 | 0.21 |
| FRW AB FC | 512 ± 1 | 613 ± 4 | 13.2 | 36 | 0.32 |

The tensile test, creep test and stress rupture test results are provided in Table 2.

From the tensile test results, it is observed that the yield strength varied from a minimum value of 482 MPa for the parent metal heat treated below the β transus and subjected to furnace cooling, to the highest exhibited by the weld in the as-welded condition (563 MPa). The weld in the as-welded condition exhibited a maximum tensile strength of 604 MPa. This can be attributed to the bond zone microstructure which exhibited a fine transformed β structure. In addition to this stress relieving treatment resulted in strengthening of the weld and near weld regions, improving high temperature tensile properties of the as-welded specimen.

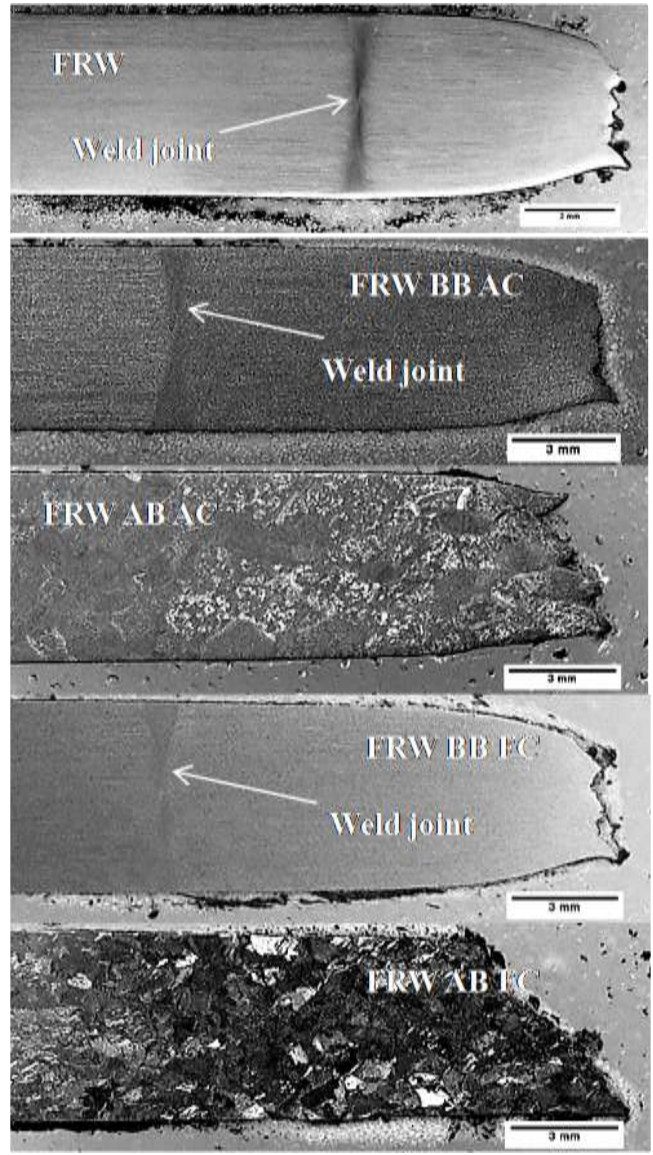

Figure 3. Longitudinal cross-sectional fractographs of the tensile tested specimens, indicating the location of failure. α+β treated specimens exhibited cup and cone fracture, whereas β treated specimens exhibited shear fracture features.

From the longitudinal cross-sectional fractographs of the tensile tested specimens (Figure 3), it is observed that all the plain tensile specimens failed in the parent metal region away from the weld, indicating that the weld region is stronger. A shift in the failure location further away from the weld region is observed after subjecting the welds to post weld heat treatments. The longitudinal sections of the tensile specimens also reveal "saw tooth" like fracture features suggesting dynamic strain ageing (DSA). The phenomenon observed is in conformity with that reported by earlier researchers that titanium alloys exhibit DSA. This behaviour is due to the pinning of dislocations by hydrogen at temperatures below 325 °C while it is due to oxygen at higher temperatures.

In general, the strength and percentage elongation reduced subsequent to post weld heat treatment. Maximum ductility is exhibited by the parent metal in the as-received condition.

The creep properties of the Ti6Al4V parent metal and welds were also evaluated at a temperature of 400 $^{o}$C.

Furnace cooling of the welds after heat treatment in the β region resulted in lower creep strain due to thick grain boundary α. Air cooling of the welds, when heat treated in the α+β region also resulted in lower creep strain. The lower creep strains observed as a result of furnace cooling is in conformity with that reported in near-α titanium alloys [10]. It can be concluded that, the parent metal and the welds heat treated in α+β and β regions followed by air cooling resulted in a significant improvement in the creep properties. The specimens subjected to β heat treatment followed by air cooling exhibited the best creep properties. The parent metal and the welds heat treated in β condition followed by furnace cooling also resulted in an improvement in the creep properties in comparison to the as-received and the as-welded conditions. But the improvement in the air cooled specimens is much better than the furnace cooled specimens. On the contrary the specimens heat treated in the α+β condition followed by furnace cooling displayed lower creep properties compared to all the other specimens.

From the data on time to rupture, in stress rupture tests presented in Table 2, it is observed that the stress rupture life varied from a minimum value of 30 h for the as-received parent subjected to β heat treatment followed by furnace cooling, and a maximum value of 212 h was recorded for the weld in the as-welded condition. Post weld heat treatment in general led to reduction in stress rupture life. Furnace cooling of the weld from above the β region resulted in lower stress rupture life of 36 h as compared to the air cooling of welds from above the β region (171 h). Furnace cooling of the welds below the β region results in lower stress rupture life compared to that for air cooled welds. This observation is similar to that in creep tests. It is observed that all the welded specimens failed outside the weld region suggesting that weld region exhibits superior creep and stress rupture properties, compared to their corresponding parent metal in respective heat treatment conditions, as the weld region contained acicular transgranular microstructure.

Coarse transgranular microstructure and coarse grains in general exhibit better creep and stress rupture properties, while finer microstructures exhibit better tensile strengths [11-12].

**CONCLUSIONS**

- The parent metal microstructure consisted of equiaxed α with β at the grain boundaries
- Heat treatment in the α+β region followed by air cooling resulted in a microstructure with primary equiaxed α plus transformed β
- Heat treatment in the β region followed by air cooling resulted in a microstructure containing widmanstatten α – β colonies
- Heat treatment in the α+β region followed by furnace cooling resulted in a microstructure of equiaxed α with intergranular β, with a discontinuous β network
- Heat treatment in the β region followed by furnace cooling resulted in coarsening of the transformed β microstructure
- Defect free Ti6Al4V friction weld joints could be obtained
- The bond zone consisted of fine transformed β structure
- Heat treatment of the welds in the α+β region followed by air cooling resulted in precipitation of β along boundaries of acicular α laths, while furnace cooling of the same resulted in coarsening of the grain boundary α
- Heat treatment of the welds in the β region followed by air cooling resulted in the formation of acicular α+β in prior β grains, whereas furnace cooling resulted in plate-like α+β in prior β grains

- Tensile properties revealed that Tensile test specimens failed in the parent metal region away from the weld, indicating weld region is stronger
- α+β treated specimens exhibited cup and cone fracture, whereas β treated specimens exhibited shear fracture
- Parent metal and the welds heat treated in α+β and β regions followed by air cooling exhibited significant improvement in the creep properties
- β heat treatment followed by air cooling exhibited the best creep properties
- The stress rupture life of the welds is better than the parent metal